\documentclass{aa}  

\usepackage{graphicx}

\newcommand{\Msun}{M$_{\odot}$}
\newcommand{\Rsun}{R$_{\odot}$}
\usepackage{txfonts}

\usepackage{color}
\usepackage{natbib,twoopt}
\usepackage{lscape} 
\usepackage{float}
\usepackage{subfig}
\usepackage[hyphenbreaks]{breakurl}
\usepackage[breaklinks]{hyperref}      
\bibpunct{(}{)}{;}{a}{}{,}             
\definecolor{cobalt}{rgb}{0.06, 0.2, 0.65}
\hypersetup{
  colorlinks,
  citecolor=blue,
  linkcolor=blue,
  urlcolor=blue,
}
\makeatletter
  \newcommandtwoopt{\citeads}[3][][]{\href{http://adsabs.harvard.edu/abs/#3}%
    {\def\hyper@linkstart##1##2{}%
     \let\hyper@linkend\@empty\citealp[#1][#2]{#3}}}
  \newcommandtwoopt{\citepads}[3][][]{\href{http://adsabs.harvard.edu/abs/#3}%
    {\def\hyper@linkstart##1##2{}%
     \let\hyper@linkend\@empty\citep[#1][#2]{#3}}}
  \newcommandtwoopt{\citetads}[3][][]{\href{http://adsabs.harvard.edu/abs/#3}%
    {\def\hyper@linkstart##1##2{}%
     \let\hyper@linkend\@empty\citet[#1][#2]{#3}}}
  \newcommandtwoopt{\citeyearads}[3][][]%
    {\href{http://adsabs.harvard.edu/abs/#3}
    {\def\hyper@linkstart##1##2{}%
     \let\hyper@linkend\@empty\citeyear[#1][#2]{#3}}}
\makeatother

\begin{document}

   \title{From Main Sequence Binary to Blast: MESA Modeling of the Double-Detonation Progenitor PTF1~J2238+7430} 
   \titlerunning{MESA Modeling of PTF1 J2238+7430}

   \author{Mercedes S. Hernandez\inst{1} \and
          Thomas Kupfer\inst{1,2} \and Diogo Belloni\inst{3} \and
          Matthias R.Schreiber\inst{4}
          }
    \authorrunning{Hernandez et al.}
      \institute{Hamburger Sternwarte, University of Hamburg, Gojenbergsweg 112, 21029 Hamburg, Germany. 
      \\\email{mercedes.hernandez.padilla@uni-hamburg.de, thomas.kupfer@uni-hamburg.de}
    \and
    Department of Physics and Astronomy, Texas Tech University, 2500 Broadway, Lubbock, TX 79409, USA. 
    \and
    International Centre of Supernovae (ICESUN), Yunnan Key Laboratory of Supernova Research, Yunnan Observatories, CAS, Kunming 650216, China
    \and
    Departamento de F{\'i}sica, Universidad T{\'ecnica} Federico Santa Mar{\'i}a, Avenida Espa{\~n}a 1680, Valpara{\'i}so, Chile. \\
}

   \date{Received September 15, 1996; accepted March 16, 1997}


\abstract
{Hot subdwarf B (sdB) stars in close binaries with white dwarf (WD) companions are  potential progenitors for double-detonation thermonuclear supernovae. The recently discovered system PTF1\,J2238+7430 is a candidate for this evolutionary channel, hosting a low-mass sdB and a comparatively massive WD in a compact orbit.}
{We aim to reproduce the evolutionary history of PTF1\,J2238+7430, in which the sdB forms first via stable mass transfer, followed by the formation of the WD through a subsequent common-envelope (CE) phase. Additionally, we seek to constrain the range of initial binary parameters that can lead to such double-detonation progenitors.}
{Using the Modules for Experiments in Stellar Astrophysics (MESA), we performed detailed binary evolution simulations from the zero-age main sequence to the present-day configuration. We explored initial stellar masses, orbital periods, and mass-loss fractions, including the effects of angular momentum transfer, tidal synchronization, and gravitational-wave-driven orbital evolution. The post-CE binary properties were derived using the standard energy formalism during CE evolution.}
{Our models successfully reproduce the observed properties of PTF1\,J2238+7430: a $0.406\,M_\odot$ sdB and a $0.72\,M_\odot$ WD in a $76.34$-minute orbit. Stable Roche-lobe overflow of a $\sim2.7\,M_\odot$ donor produces the sdB, while the WD forms from the initially less massive companion during an episode of CE evolution. We find that the CE ejection efficiency must be high ($\alpha_{\rm CE} \approx 0.87$) to match the observed orbit, exceeding canonical values for similar systems. We further delineate the allowed parameter space for initial binaries that can evolve into sdB+WD systems consistent with double-detonation progenitors. These limits are preliminary; a systematic exploration of all parameters is needed for robust constraints, we highlight the main challenges in our MESA simulations and provide a useful starting point for future work.
}
{Our findings identify promising regions of the parameter space for forming PTF1\,J2238+7430–like systems and provide a foundation for future systematic studies of sdB+WD binaries as potential double-detonation Type Ia supernova progenitors.  
}


 \maketitle
%

\section{Introduction}

Most hot subdwarf B stars are the stripped cores of red giants that continue to burn helium in their centers while retaining only a thin hydrogen envelope. With typical masses of about $0.5\,M_\odot$, they represent an advanced evolutionary stage and are most commonly found in close binaries \citep{Heber1986, Heber2009, Heber2016, Ulrich2026}. Many sdB stars have orbital companions with periods shorter than $10$ days \citep{Maxted2001, Napiwotzki2004, schaffenroth2022}, and in some extreme cases, orbital periods below one hour have been observed \citep[e.g.,][]{Vennes2012, Geier2013, Kupfer2017a, Kupfer2017b, Kupfer2020a, Kupfer2020b}. These compact systems are generally thought to form through a common envelope interaction, which drastically reduces orbital separation. If the post-CE orbital period is shorter than about two hours, the sdB is expected to fill its Roche lobe while still undergoing core helium burning \citep{Bauer2021}. Gravitational radiation drives the system toward tighter orbits, typically at orbital periods between $30$ and $100$ minutes \citep[e.g.,][]{Savonije1986, Tutukov1989, Tutukov1990, Iben1991, Yungelson2008, Piersanti2014, Brooks2015, Neunteufel2019, Bauer2021}.

These binaries are of particular astrophysical significance because they provide potential progenitors for thermonuclear explosions through the double-detonation channel. 
In this scenario, the sdB transfers helium-rich matter onto the surface of its white dwarf companion. If the accumulated helium layer reaches a critical mass, it can ignite unstably. This ignition may either trigger a secondary detonation in the white dwarf core, producing a classical double-detonation supernova even below the Chandrasekhar limit \citep[e.g.,][]{Livne1990, Livne1995, Fink2010, Woosley2011, Wang2012, Shen2014, Wang2018,Rajamuthukumar2024}, or result only in a surface helium detonation, giving rise to a faint and rapidly evolving supernova Ia event, possibly followed by weaker helium flashes \citep{Bildsten2007, Brooks2015}.  

From an evolutionary perspective, sdB stars are generally believed to form through three distinct channels. The second common-envelope (CE) ejection channel is the primary mechanism responsible for most of the observed compact sdB+WD systems. In this scenario, the sdB progenitor—already orbiting a white dwarf—undergoes a CE phase, and the envelope ejection leaves behind a very close binary with short orbital periods ranging from about 0.5 hours to 25 days. The second channel corresponds to stable Roche-lobe overflow (RLOF), which involves stable mass transfer onto a WD companion. Although this channel is considered less likely, as it requires a relatively massive WD companion, it would lead to wide binaries with long orbital periods, typically around 1000 days. Finally, it is important to distinguish the third main formation channel—the merger of two helium white dwarfs—which results in a single sdB star rather than a binary system. In all cases, the white dwarf is formed first during a CE episode, while the sdB emerges later through different evolutionary pathways \citep{Han2002, Han2003}. However, recent observations suggest that these sequences may not always hold. Two systems have been identified in which the sdB appears to have formed prior to the white dwarf: CD--30$^\circ$11223, with a $70.5$ minutes orbital period and a massive white dwarf companion of about $0.75\,M_\odot$ \citep{Vennes2012, Geier2013, deshmukh2024}, and PTF1~J2238+7430, a close binary with a $76$ minutes orbital period that has been proposed as a candidate progenitor of a thermonuclear double-detonation supernova \citep{Kupfer22}.  

\citet{Kupfer22} predict the future evolution of PTF1~J2238+7430. According to their calculations, the sdB component is expected to initiate mass transfer of its hydrogen-rich envelope in about $6$ Myr, at a relatively low rate \citep{Bauer2021}. After $\sim 60$ Myr, while the sdB is still undergoing core-helium burning, it is predicted to start transferring helium-rich material to the white dwarf companion. This mass transfer episode is expected to lead to the accumulation of a substantial helium layer of $\sim 0.17\,M_\odot$, raising the total white dwarf mass to $0.92\,M_\odot$. At this stage, the models predict that the white dwarf will undergo a thermonuclear instability, triggering a detonation that is likely to disrupt the star in a peculiar thermonuclear supernova \citep{Woosley2011, Bauer2017}. However, recently \citet{piersanti2024} and \citet{Rajamuthukumar2024} suggested that PTF1~J2238+7430 only shows several He-flashes and a thermonuclear supernova is prevented.

PTF1\,J2238+7430 is particularly remarkable. Although it is a single-lined spectroscopic binary, detailed modeling shows that it consists of a low-mass hot subdwarf star ($M_{\mathrm{sdB}} = 0.383 \pm 0.028\,M_\odot$) and a comparatively massive white dwarf companion ($M_{\mathrm{WD}} = 0.725 \pm 0.026\,M_\odot$), corresponding to a mass ratio of $q = 0.528 \pm 0.020$ \citep{Kupfer22}. The binary is nearly edge-on ($i \approx 88.4^\circ$) and displays weak eclipses of the white dwarf, with an orbital period of $P_{\mathrm{orb}} = 76.34179(2)$ minutes. By modelling the eclipses \citet{Kupfer22} were able to estimate the temperature of the white dwarf to be $26,800 \pm 4600$ K and a radius of $0.0109\,R_\odot$, consistent with theoretical models of carbon–oxygen white dwarfs \citep{Romero2019}. The sdB star rotates at $185 \pm 5$ km~s$^{-1}$, consistent with tidal synchronization \citep{Kupfer22}.  

A notable feature of this system is the discrepancy between the evolutionary ages of its two components: the cooling age of the white dwarf is estimated at $\sim 25$ Myr, whereas the sdB age is predicted to be about $\sim 170$ Myr \citep{Kupfer22}. This tension can be resolved in a formation scenario where the sdB forms first via stable mass transfer, and the white dwarf companion was formed second, following a CE episode about 25 Myr ago \citep{Ruiter2010}. \citet{Kupfer22} did not perform a detailed modeling of the evolutionary history. They just discussed that the progenitor of sdB was likely a $\sim 2\,M_\odot$ main-sequence star based on the sdB mass.  
\\

 In this work, we aim to reproduce the evolutionary pathway proposed by \citet{Kupfer22} and \citet{Ruiter2010}, in which the sdB star forms first through a phase of stable mass transfer, followed by the formation of the white dwarf companion during a subsequent common-envelope episode. Using detailed simulations with Modules for Experiments in Stellar Astrophysics MESA, we trace the system’s evolution from the main sequence to its present configuration. Beyond reproducing this specific case, our goal is to constrain the parameter space in which progenitors of double-detonation supernovae may arise on the main sequence. In doing so, we also explore the implications of our models for the efficiency of the common-envelope phase, a key but still poorly constrained process in binary stellar evolution \citep[]{zorotovic10-1,Zorotovic22}

\section{MESA Simulations}

 Our aim is to reproduce the origin of PTF1\,J2238+7430 using the Modules for Experiments in Stellar Astrophysics tool \citep[version r24.08.1 of the MESA code][]{Paxton2011, Paxton2013, Paxton2015, Paxton2018, Paxton2019, Jermyn2023}, following the evolutionary framework proposed by \citet{Kupfer22}. 
 or the MESA simulations, the equation of state combines several sources: OPAL \citep{Rogers2002}, SCVH \citep{Saumon1995}, FreeEOS \citep{Irwin12}, HELM \citep{Timmes2000}, PC \citep{Potekhin2010}, and Skye \citep{Jermyn2021}. Nuclear reaction rates are drawn from a mixture of NACRE \citep{Angulo1999}, JINA REACLIB \citep{Cyburt2010}, along with additional tabulated weak reaction rates \citep{Fuller1985,Oda1994,Langanke2000}. Screening effects are incorporated following the prescription of \citet{Chugunov2007}, and thermal neutrino losses are based on \citet{Itoh1996}. Electron conduction opacities are taken from \citet{Cassisi2007}, while radiative opacities are primarily sourced from OPAL \citep{Iglesias1993,Iglesias1996}, with high-temperature, Compton-scattering-dominated conditions calculated using the formulae of \citet{Buchler1976}.

\subsection{ sdB formation through stable mass transfer}

\begin{figure}
    \centering
    \includegraphics[width=\linewidth]{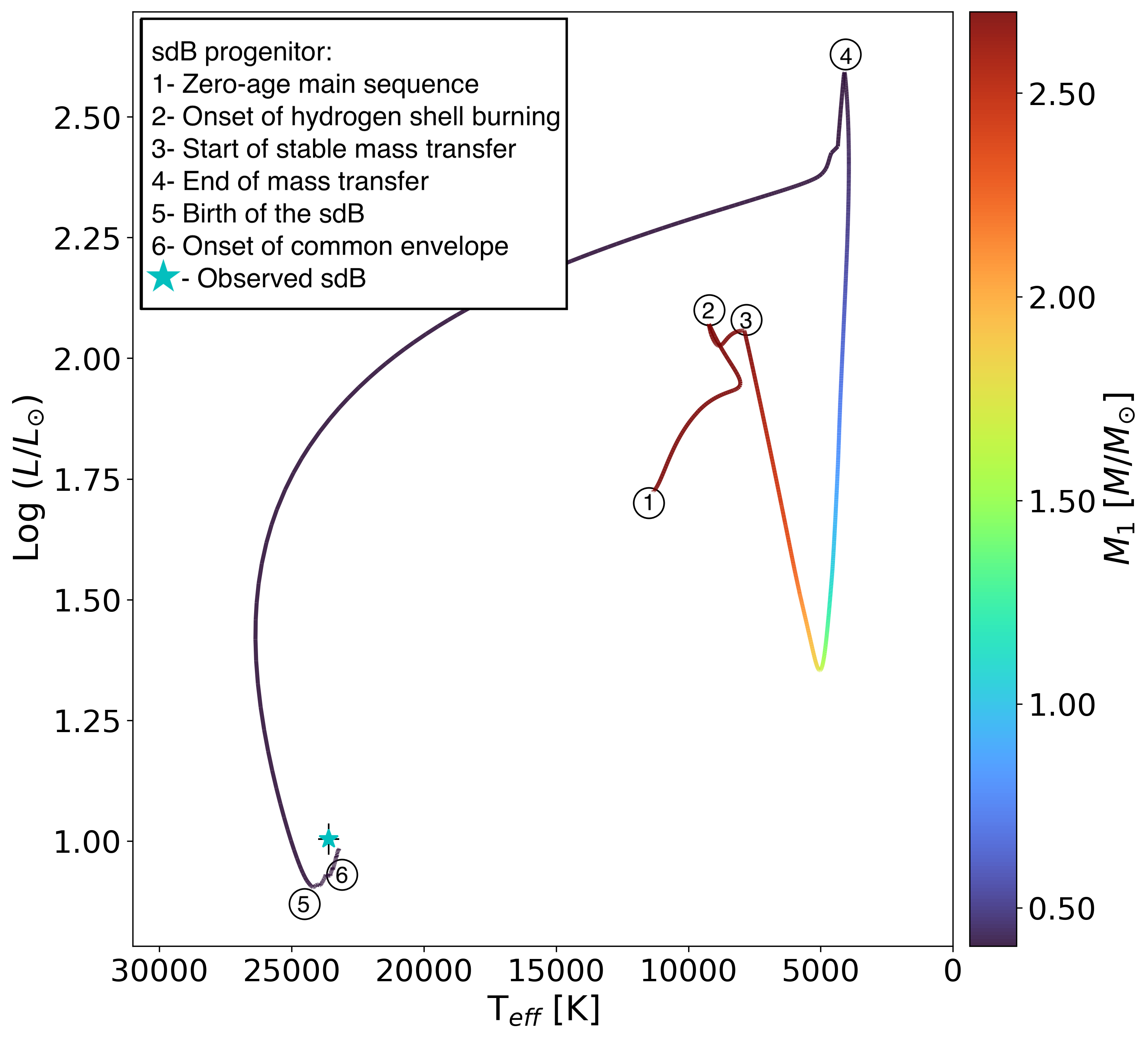}
    \caption{
    MESA evolutionary track of the sdB progenitor in a binary system with initial stellar masses of $2.7\,M_\odot$ (primary) and $2.6\,M_\odot$ (secondary), and an initial orbital period of 3 days. Point “1” marks the start of the simulation at the zero-age main sequence. Point “2” corresponds to the onset of hydrogen shell burning, while circle “3” indicates the beginning of stable mass transfer, which terminates at point “4.” At point “5,” the stripped primary becomes an sdB star. Finally, point “6” marks the onset of the common-envelope phase initiated by the secondary. The color of the track represents the mass evolution of the primary star. The blue star indicates the observed position of the sdB component of PTF1~J2238+7430, as reported by \citet{Kupfer22}.
    }
    \label{fig:HRD}
\end{figure}

\begin{figure}
    \centering
    \includegraphics[width=\linewidth]{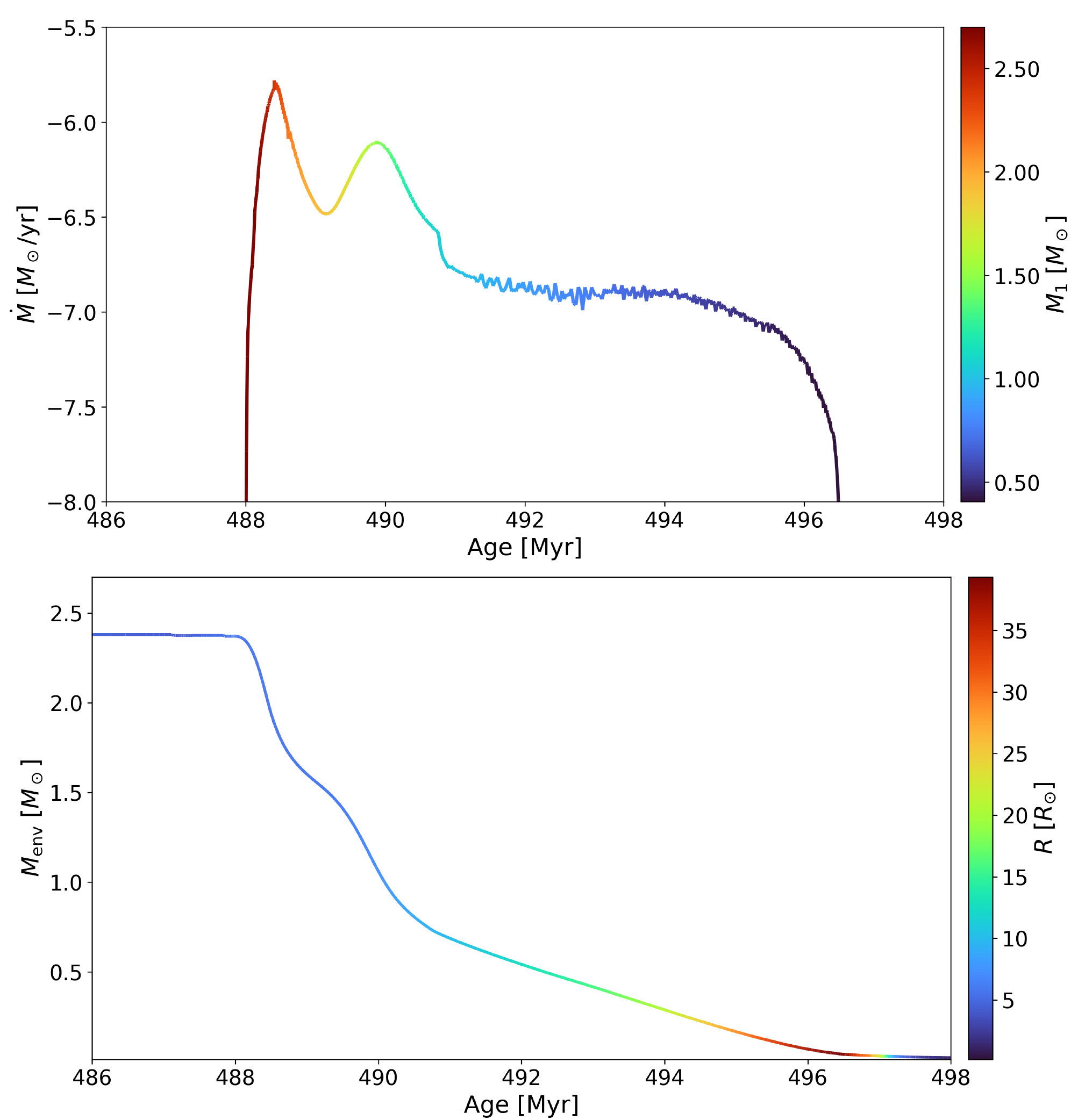}
    \caption{
    Evolution of the $2.7\,M_\odot$ main-sequence star during the stable mass-transfer episode. 
    \textit{Top panel:} Mass-transfer rate, with colors indicating the change in mass of the sdB progenitor.  
    \textit{Bottom panel:} Stripping of the H-rich envelope, with colors indicating the stellar radius as the star evolves to reveal its core (sdB).
    }
    \label{fig:MT-Hshell}
\end{figure}

The evolutionary scenario for PTF1\,J2238+7430 proposed by \citet{Kupfer22} suggests that the sdB star forms first through stable mass transfer. To ensure this outcome, the initial masses of the main-sequence stars were chosen based on the adiabatic response of the donor star. This stability condition is typically satisfied when the initial mass ratio, $q_i$, is below a critical threshold, $q_{\mathrm{crit}}$. For conservative mass transfer involving donors with masses between 1 and 6\,\Msun, $q_{\mathrm{crit}}$ at the end of the red giant branch phase is estimated to lie between 0.7 and 1 \citep{Ge2020}.

We perform a series of MESA simulations that evolve both components of a binary system initialized at the zero-age main sequence (ZAMS), where the initial conditions range from $1.8$ to $3.5\,M_\odot$ for the sdB progenitor \citep[ Figure 4]{Arancibia-Rojas2024}, $1.7$ to $3.0\,M_\odot$ for the companion, and orbital periods between $1.0$ and $50$ days. 
We investigate different values of the mass-loss fraction during stable mass transfer, $\beta$, ranging from $0.85$ to $0.15$ \citep{Lechien2025}. In this prescription, the companion star accretes between 15\% and 85\% of the material lost from the donor’s envelope.

The simulations are performed using the MESA binary module under the following assumptions. The mass-transfer rate is computed using the Ritter scheme, allowing for stable Roche-lobe overflow. We adopt non-conservative mass transfer with $\beta>0$, where $\beta$ represents the fraction of transferred material that is accreted by the companion star. The parameters $\alpha$, $\delta$, and $\gamma$, which describe different modes of angular momentum loss — respectively, through material leaving the vicinity of the accretor, through a circumbinary disk, and through isotropic re-emission or other system-wide outflows — are all set to zero. Angular momentum losses are included from gravitational wave radiation, systemic mass loss (i.e. mass escaping the binary system entirely and carrying angular momentum away), and spin–orbit coupling, while magnetic braking is disabled.
Tidal circularization and synchronization are enabled, with both stars treated as rotating rigid bodies. Wind mass loss and accretion are allowed for both components, with a Bondi–Hoyle accretion efficiency of 1.5 and a maximum capture fraction of 1.0.

The stellar structure is evolved at solar metallicity ($Z=0.02$) using a nuclear network that expands adaptively as advanced burning stages are encountered. Convection is treated via mixing-length theory with $\alpha_{\rm MLT}=2.0$ (Henyey formulation), applying the Ledoux criterion together with semi-convection (efficiency = 1.0) and thermohaline mixing (coefficient = 2.0). We also included predictive mixing as described in \citet{Ostrowski_2021} as well as 
rotationally induced mixing processes (Solberg-Hoiland, secular shear instability, Eddington-Sweet circulation, Goldreich-Schubert-Fricke and Spruit-Tayler dynamo) following \citet{Heger_2000,Heger_2005}.
Convective core overshooting is included using the exponential prescription with $f=0.016$ and $f_0=0.008$ for $M > 2\,M_\odot$.
Mass loss on the RGB and AGB is modeled with Reimers ($\eta=0.1$) and Blöcker ($\eta=0.02$) prescriptions, respectively. To ensure numerical stability during rapid evolutionary phases, such as the helium flash, restrictive timestep and convergence controls are applied.
Stellar rotation is initialized at 1\% of the critical rate for both stars, and the accretor is allowed to accrete angular momentum and spin up during stable mass transfer.
However, if the breakup limit is reached, we assume that rotation acts as a regulating mechanism for the mass transfer rate \citep[e.g.][]{Piersanti2003}; that is, the accretor is not allowed to spin up beyond the breakup limit.

We found that the parameters required to successfully reproduce the properties of PTF1~J2238+743 are the initial stellar masses of $2.70\,M_\odot$ (donor) and $2.60\,M_\odot$ (accretor), with an initial orbital period of $3.0$~days and adopting a mass-loss fraction of $\beta = 0.15$, meaning that 15\% of the transferred material is lost from the vicinity of the accretor, carrying its specific angular momentum. The simulation models a close binary system starting with both stars on the main sequence. The system undergoes a phase of stable Roche-lobe overflow, during which mass is transferred from the initially more massive donor to the secondary. As a result, the donor is stripped of most of its envelope and becomes an sdB star, while the accretor gains mass and eventually becomes a white dwarf.

Figure~\ref{fig:HRD} shows the formation of the sdB star through stable mass transfer, as obtained from our MESA simulation, where the $2.7,M_\odot$ primary is the sdB progenitor. The evolutionary track starts at point~1, when both stars are on the main sequence.
After $\sim487$~Myr, the primary evolves into a subgiant, developing a helium core mass of $0.323,M_\odot$ (point~2). Shortly afterwards, at an orbital period of $2.9$~days, the primary fills its Roche lobe and initiates stable, non-conservative mass transfer (point~3).
Mass transfer initially proceeds on the thermal timescale of the donor, lasting $\sim2$~Myr with rates of order $\dot{M} \sim 10^{-6},M_\odot,\mathrm{yr}^{-1}$, during which the donor loses about $1.0$--$1.2,M_\odot$ (top panel Figure,\ref{fig:MT-Hshell}). Subsequently, the system enters a longer phase of mass transfer on the nuclear timescale of the donor, at a reduced rate of $\dot{M} \sim 10^{-7},M_\odot,\mathrm{yr}^{-1}$.
Roche-lobe overflow ends after $\approx 9$~Myr, at an age of $496.6$~Myr. By this time, the envelope of the primary has been nearly stripped away (bottom panel Figure,\ref{fig:MT-Hshell}), leaving behind a proto-sdB star of $0.406,M_\odot$ in a binary with an orbital period of $157.9$~days (Figure~\ref{fig:HRD} point~4). Helium ignition occurs shortly afterwards, marking the transformation into a core helium-burning sdB star with a residual hydrogen envelope of $M_{\rm env}= 0.0137 \,M_\odot$ (Figure~\ref{fig:HRD} point~5 and Figure~\ref{fig:MT-Hshell} bottom). Table~\ref{Tab:FormationChannel} summarizes all the evolutionary details.

\subsection{ White dwarf formation through common envelope}

\begin{figure*}
    \centering
    \includegraphics[width=1.0\linewidth]{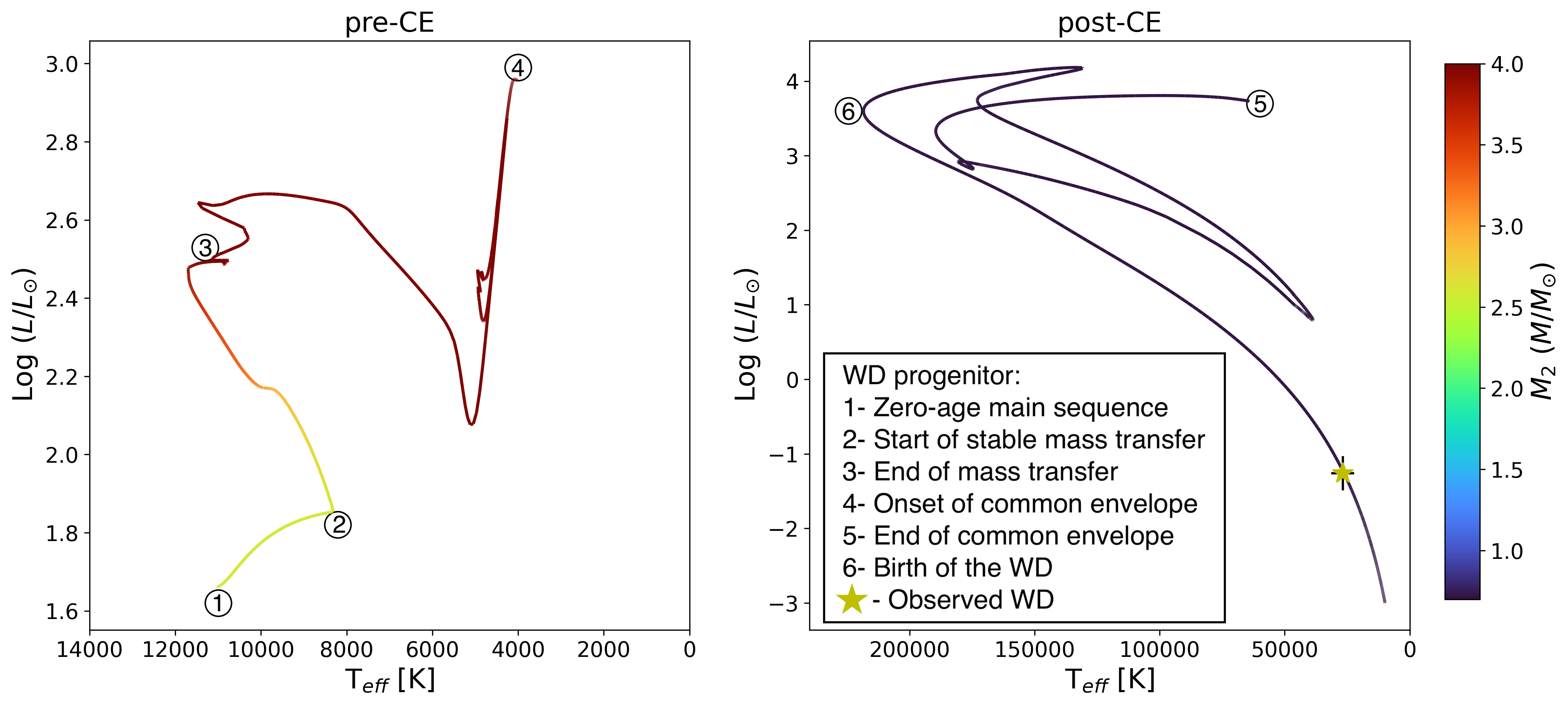}
    \caption{Evolutionary track of the $2.6,M_\odot$ star, i.e., the white dwarf progenitor, as obtained from two MESA simulations. The left panel, which corresponds to the same simulation shown in Figure~\ref{fig:HRD}, illustrates the pre-common-envelope evolution, starting at the zero-age main sequence, followed by the onset and termination of stable mass transfer, and culminating in the initiation of the common-envelope phase. The right panel shows the post-common-envelope evolution: the white dwarf cooling track was simulated by removing the red giant envelope, the system emerges as a compact binary, the secondary contracts to form a white dwarf, and the track continues until it reaches the present-day configuration of PTF1~J2238+7430, marked with a yellow star. The color bar indicates the change in the mass of the WD progenitor in both panels.}
    \label{fig:HR2}
\end{figure*}

\begin{figure}
    \centering
    \includegraphics[width=0.9\linewidth]{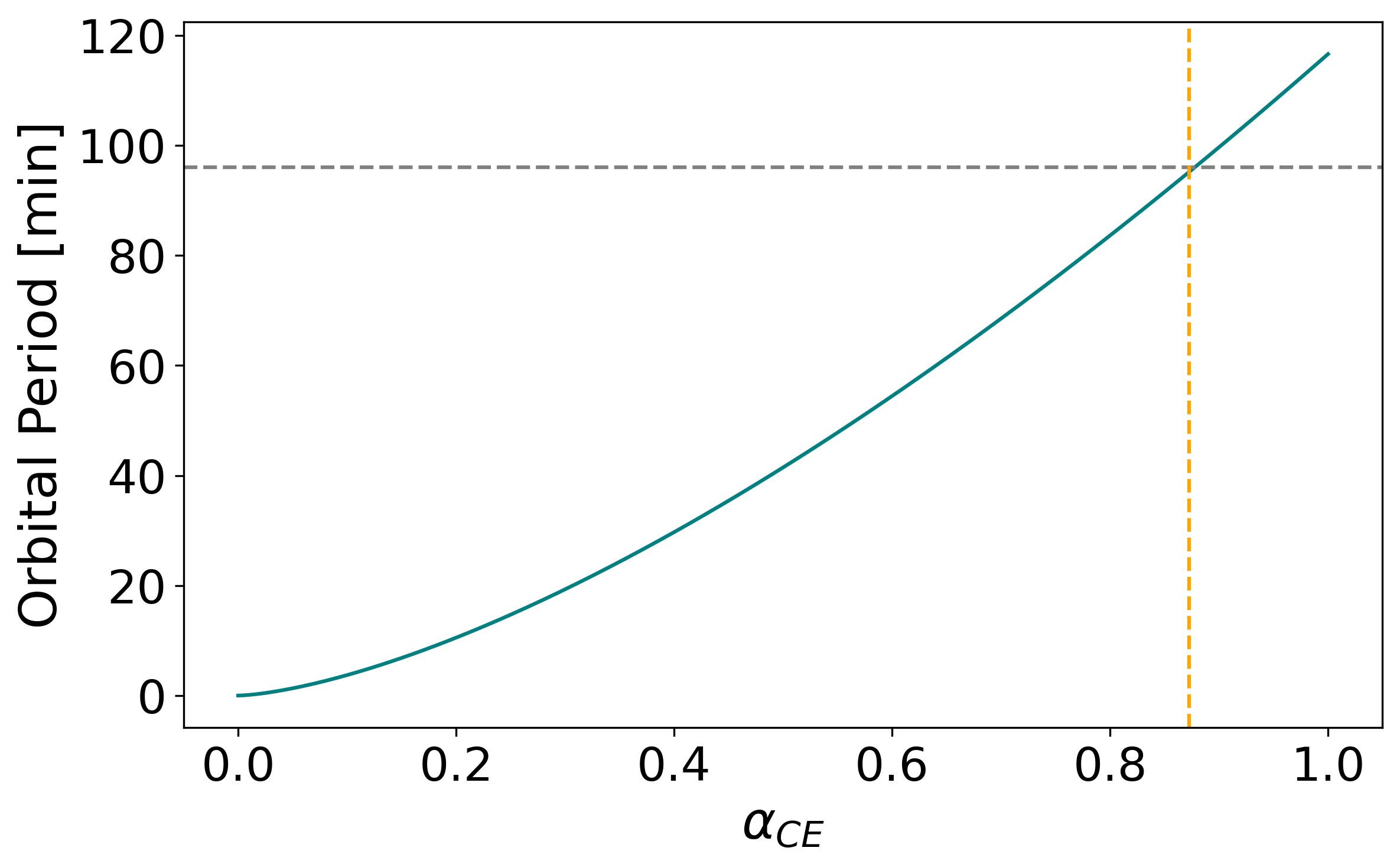}
    \caption{Common-envelope efficiency ($\alpha_{\mathrm{CE}}$) versus final orbital period. The solid line shows the possible post–common-envelope orbital configurations. The horizontal dashed line indicates the predicted orbital period of $95.04$ minutes for PTF1\,J2238+7430 \citep{Kupfer22}, while the vertical orange dashed line marks the corresponding $\alpha_{\mathrm{CE}}=0.872$. The recombination energy was fixed to $\lambda=1$.
}
    \label{fig:CEE}
\end{figure}

The evolution of the white dwarf progenitor is illustrated in Figure~\ref{fig:HR2}. In the left panel, the system begins as a zero-age main-sequence binary (point 1). At $487.96$~Myr, the secondary component (future WD) is still on the main sequence when the primary fills its Roche lobe, and stable mass transfer starts (point 2). This episode lasts until $496.61$~Myr, at which time the primary has been stripped to a $0.406\,M_\odot$ proto-sdB star and the secondary has accreted mass, reaching $3.99\,M_\odot$ (point 3).

After the end of stable mass transfer, the secondary continues to evolve. At $530.6$~Myr it leaves the main sequence and soon after ascends the giant branch, igniting helium in its core. During this phase, the star develops an extended envelope, and tidal interactions become significant. The tidal forces act to synchronize the red giant’s rotation with the orbital motion, resulting in a gradual reduction of the orbital separation and period (from about 153.264 to 61.839 days) as the star evolves.  
By $580.1$~Myr, the secondary reaches the asymptotic giant branch (AGB). At $582.13$~Myr, it expands enough to fill its Roche lobe, initiating a second mass-transfer phase that rapidly becomes unstable and triggers a common-envelope (CE) episode (point 4).

The right panel of Figure~\ref{fig:HR2} shows the binary at the end of the CE phase (point~5), when it emerges with an orbital period of $95$~minutes and a $0.72\,M_\odot$ proto-WD. By an age of $583.6$~Myr, the secondary has contracted and becomes a white dwarf (point~6). To follow this post-CE evolution, we performed a second MESA run starting from this configuration. In this calculation, we explored different initial orbital periods after envelope ejection and, in parallel, evolved the white dwarf on the cooling track by removing the red giant envelope. By adjusting the post-CE orbital period, we obtained a configuration in which, after $\sim 50$~Myr of binary evolution, both the orbital period and the WD effective temperature match the observed properties of PTF1~J2238+7430. At $632.5$~Myr, the binary indeed resembles the observed system (marked with a yellow star in Figure~3), consisting of a $0.405,M_\odot$ sdB and a $0.722,M_\odot$ WD in a $76.3$-minute orbit.  The rotational state of the components after the common-envelope phase is uncertain; therefore, we do not assume any initial rotation in the post-CE sdB+WD model.

Now that the configuration of the system both before and after the common envelope phase is known, we can calculate the common envelope efficiency using the standard energy formalism of \citet{Webbink1984}.

We calculated the envelope binding energy ($E_{\mathrm{bind}}$) by integrating the star envelope from the helium core boundary (i.e., at a radius at which the helium mass fraction is 0.1) to the surface of the star, that is,

\begin{equation}
E_{\rm bind} \ = \ 
- \int_{M_{\rm d,c}}^{M_{\rm d}} \frac{G \; m}{r(m)} \; {\rm d}m 
 \ + \  
\int_{M_{\rm d,c}}^{M_{\rm d}} \varepsilon_{\rm int}(m) \; {\rm d}m \ ,
\label{Eq:AlphaINT}
\end{equation}

\noindent
where the first integral is the potential energy ($E_{\mathrm{grav}}$), the second integral is the internal energy ($E_{\mathrm{int}}$),  $r$ is the radius, $m$ is the mass, $\varepsilon_{\rm int}$ is the specific thermodynamic internal energy.
Our implementation to compute the thermodynamic internal energy is the same as in \citet{Belloni_2024c,Belloni_2024d}.
At the onset of CE evolution, $E_{\mathrm{grav}} = -1.4641\times10^{48}\,\mathrm{erg}$ and $E_{\mathrm{int}} = 7.9228\times10^{47}\,\mathrm{erg}$.

The initial orbital energy is computed from
\begin{equation}
E_{\mathrm{orb,i}} = \frac{G\,M_{\mathrm{c,1}}\,M_{\mathrm{c,2}}}{2\,a_{\mathrm{i}}},
\end{equation}
where $M_{\mathrm{c,1}} = 0.7217\,M_\odot$ is the CO core mass of the primary at the onset of the CE,   
$M_{\mathrm{c,2}} = 0.4054\,M_\odot$ is the total mass of the secondary,   
and $a_{\mathrm{i}} = 107.726\,R_\odot$ is the binary separation.   
For these values, we obtain $E_{\mathrm{orb,i}} \approx 3.521\times10^{46}\,\mathrm{erg}$.

The final orbital energy is given by
\begin{equation}
E_{\mathrm{orb,f}} = E_{\mathrm{orb,i}} - \frac{E_{\mathrm{bind}}}{\alpha_{\mathrm{CE}}},
\end{equation}
where $\alpha_{\mathrm{CE}}$ is the common-envelope ejection efficiency.  
For each assumed $\alpha_{\mathrm{CE}}$ in the range $0.0 \leq \alpha_{\mathrm{CE}} \leq 1.0$, we calculate the corresponding final orbital separation
\begin{equation}
a_{\mathrm{f}} = \frac{G\,M_{\mathrm{c,1}}\,M_{\mathrm{c,2}}}{2\,E_{\mathrm{orb,f}}},
\end{equation}
and derive the post–CE orbital period from Kepler’s third law. This procedure allows us to map the possible final orbital configurations for the given pre– and post–CE parameters (see Figure~\ref{fig:CEE}), where the post–CE orbital period is shown as a function of the CE ejection efficiency. For PTF1\,J2238+7430, the observationally inferred post–CE period ($95$~min) corresponds to $\alpha_{\mathrm{CE}} \approx 0.87$. A detailed comparison between the observed properties of PTF1~J2238+7430 and the outcomes of our MESA simulations is presented in Table~\ref{tab:final}.

\begin{table*}
\centering
\caption{Evolution of a zero-age main-sequence binary towards PTF1~J2238+7430.}
\label{Tab:FormationChannel}
\setlength\tabcolsep{6pt} 
\renewcommand{\arraystretch}{1} 
\begin{tabular}{r c c r r c c r l}
\hline
\noalign{\smallskip}
\multicolumn{1}{c}{Time}   &   $M_1$    &   $M_2$   &  \multicolumn{1}{c}{$R_1$}    &  \multicolumn{1}{c}{$R_2$}    & Type$_1$  & Type$_2$  & \multicolumn{1}{c}{$P_{\rm orb}$} & Event\\
\multicolumn{1}{c}{(Myr)}  & (M$_\odot$)&(M$_\odot$)&\multicolumn{1}{c}{(R$_\odot$)}&\multicolumn{1}{c}{(R$_\odot$)}&           &           &  \multicolumn{1}{c}{(days)}       &      \\
\hline
\noalign{\smallskip}
     0.000  &  2.700  &  2.600  &   1.897  &   1.859       & MS         & MS       &   3.000  &  zero-age MS binary     \\
   487.137  &  2.699  &  2.599  &   4.233  &   4.063       & SG         & MS       &   2.926  &  change in primary type \\
   487.964  &  2.699  &  2.599  &   5.668  &   4.076       & SG         & MS       &   2.913  &  onset of stable mass transfer \\
   489.552  &  1.713  &  3.437  &   6.250  &   3.938       & FGB        & MS       &   4.391  &  change in primary type \\
   496.502  &  0.406  &  3.991  &  39.137  &   5.058       & proto-sdB  & MS       & 157.590  &  change in primary type (He ignition) \\
   496.612  &  0.406  &  3.991  &  35.228  &   5.061       & proto-sdB  & MS       & 157.863  &  end of stable mass transfer \\
   529.883  &  0.405  &  3.990  &   0.161  &   5.850       & sdB        & MS       & 158.512  &  change in primary type \\
   530.585  &  0.405  &  3.988  &   0.161  &   5.610       & sdB        & SG       & 158.606  &  change in secondary type \\
   533.074  &  0.405  &  3.984  &   0.162  &  13.852       & sdB        & FGB      & 158.904  &  change in secondary type \\
   533.438  &  0.405  &  3.984  &   0.162  &  48.586       & sdB        & CHeB     & 152.609  &  change in secondary type \\
   580.092  &  0.405  &  3.981  &   0.172  &  30.736       & sdB        & AGB      & 153.264  &  change in secondary type \\
   582.134  &  0.405  &  3.981  &   0.173  &  62.158       & sdB        & AGB      &  61.839  &  onset of CE evolution\\
   582.134  &  0.405  &  0.722  &   0.173  &   0.594       & sdB        & proto-WD &  0.066  &  end of CE evolution \\
   583.569  &  0.405  &  0.722  &   0.175  &   0.044       & sdB        & WD       &   0.065  &  change in secondary type \\
632.480  &  0.405  &  0.722  &  0.192  &  0.011       & sdB     & WD &   0.053  &  binary looks like PTF1 J2238+7430 \\

\noalign{\smallskip}
\hline
\end{tabular}
\tablefoot{{MESA evolutions for the pre-CE and post-CE evolution. 
The terms $M_1$ and $M_2$, $R_1$ and $R_2$, and Type$_1$ and Type$_2$ are the masses, radii, and stellar types of the primary and secondary, respectively. $P_{\rm orb}$ is the orbital period and the last column corresponds to the event occurring to the binary at the given time in the first column. The row in which the binary has the present-day properties is highlighted in boldface. 
Abbreviations:
MS~(main~sequence~star),
SG~(subgiant~star),
FGB~(first giant branch star),
CHeB~(core helium burning),
AGB~(asymptotic giant branch star),
WD~(white~dwarf),
sdB~(hot~subdwarf~B~star),
CE~(common~envelope).
}}
\end{table*}

\begin{table}[]
    \centering
    \caption{Observational Constraints vs. MESA Models for PTF1~J2238+7430.}

    \begin{tabular}{rll}
    \hline
        Parameter & Observed & Simulation  \\
        \hline
        $P_{\rm{orb}}$ [min] & 76.34179(2) & 76.330 \\
        $M_{\rm{sdB}}$ [\Msun] & 0.383$\pm$ 0.028 & 0.405 \\
        $M_{\rm{WD}}$ [\Msun] & 0.725$\pm$0.026 & 0.722\\ 
        $R_{\rm{sdB}}$ [\Rsun] & 0.190 $\pm$0.003 & 0.192 \\
        $R_{\rm{WD}}$ [\Rsun] & 0.0109$\pm$0.0003 & 0.011\\
        $T_{\rm{eff,sdB}}$ [K] & 23,600$\pm$400 &  23,190   \\
        $T_{\rm{eff,WD}}$ [K] & 26,800$\pm$4,600 & 24,045   \\
        $log g_{\rm{sdB}}$ &5.42$\pm$0.06 & 5.477 \\
        $log g_{\rm{WD}}$ & - & 8.213 \\
        Age$_{\rm{sdB}}$ [Myr] & 170 & 104.316 \\
        Age$_{\rm{WD}}$ [Myr] & 25 & 50.35\\
        \hline\\
     \end{tabular}
     \tablefoot{{ Comparison between the observational constraints of PTF1~J2238+7430 reported by \citet[][KT22]{Kupfer22} and the results of our MESA simulations. The table lists the main stellar and orbital parameters used as benchmarks.
    }}
    \label{tab:final}
\end{table}

\section{Discussion}

The solution we obtain for the formation of the observed sdB+WD system represents only one of several possible configurations. Its validity depends sensitively on the initial assumptions adopted in the simulations. In this sense, it is essential to emphasize that our results should be regarded as illustrative rather than definitive.

\subsection{Mass and angular momentum accretion efficiencies}

The formation of the sdB+WD binary relies on several key assumptions regarding the physics of mass transfer, each of which introduces uncertainties that may affect the outcome. 
Although there is no universal prescription for the fraction of mass lost during Roche-lobe overflow episodes \citep{Webbink2008}, in our simulations we adopted a non-conservative mass transfer scheme. 
Our exploration of the mass loss fraction ($\beta$ parameter) was sparse, yet we found that systems capable of reproducing the observed configuration required highly conservative mass transfer (at least $\sim 85\%$ of the transferred mass accreted by the companion). 

This finding is broadly consistent with the results of \citet{Lechien2025}. 
In their sample of 16 stripped stars with rapidly rotating companions (Be+sdOB systems), they found that about half of the binaries required the accretor to retain at least 50\% of the transferred mass. Moreover, they showed that assuming a constant accretion efficiency--where all secondaries accrete between 60\% and 80\% of the transferred mass—-could, in principle, reproduce the entire sample. However, when analyzing individual systems with configurations similar to PTF1~J2238+7430 (such as 7~Vul and $\kappa$~Dra), reproducing the charactiristics of the system derived from observations through stable mass transfer can be achieved with substantially lower efficiencies.
This could indicate that, while our adopted assumptions allow us to reproduce the observed system, alternative solutions with different mass loss fractions may also be viable.

In our modeling, the accretion of matter and angular momentum is drastically reduced as soon as the main-sequence secondary reaches the breakup spin; that is, it accretes only the amount required to maintain rotation at the breakup limit.
In principle, limiting angular momentum accretion during mass transfer can be justified, since tidal forces eventually synchronize the accretor’s rotation with the orbit after mass transfer.
However, if matter continues to be accreted after the accretor reaches the breakup velocity, with the excess angular momentum being returned to the accretion disk as suggested by \citet{Paczynski1991} and \citet{Popham1991}, we believe that a reasonable model for PTF1~J2238+7430 could still be obtained, provided that the initial masses and orbital period are different.
That being said, exploring different assumptions for angular momentum transport in future simulations will be useful for further testing the robustness of our results.

To assess the impact of rotation on the system properties, we performed an additional simulation with the same initial parameters and mass-transfer efficiency as our best-fitting model, but neglecting rotational effects throughout the evolution. We find that the resulting sdB star is slightly cooler, slightly more luminous, slightly larger, and retains a marginally thicker hydrogen envelope. However, these differences remain within the observational uncertainties, indicating that rotation has only a minor effect on the final system properties in this case.

\subsection{Common envelope efficiency assumptions}

Our main objective was to identify a binary configuration that, upon evolution, could reproduce the observed properties of PTF1\,J2238+7430. The guiding hypothesis was that such a configuration, after undergoing common-envelope evolution, would require an efficiency parameter of $\alpha_{\mathrm{CE}} \simeq 0.2$--$0.4$, based on values inferred for post-common-envelope binaries hosting AFGKM + white dwarf companions \citep{zorotovic10-1,Hernandez2021,Hernandez2022-2,Hernandez2022,Zorotovic22}. 

However, we soon realized that achieving such low values of $\alpha_{\mathrm{CE}}$ requires that the binary, immediately after the phase of stable mass transfer, retains an orbital period of at least $\sim 400$\,days. In contrast, across all the models explored, the longest orbital period obtained after stable mass transfer was only $\sim 170$\,days, significantly below the required minimum. This discrepancy suggests that the common-envelope efficiency in the case of PTF1\,J2238+7430 must have been substantially higher than the canonical range typically adopted for other post–common-envelope systems, assuming full recombination energy efficiency ($\lambda = 1$). This result is consistent with the findings of \citet{Zhang2024}, who reported that common-envelope efficiencies tend to be higher for systems with more massive white dwarf progenitors.


Alternatively, the discrepancies in the derived values of $\alpha_{\mathrm{CE}}$ between PTF1\,J2238+7430 and systems containing AFGKM+white dwarf companions may stem from fundamental differences in their evolutionary histories. Most empirical constraints on the common-envelope efficiency parameter are based on binaries in which the common-envelope phase corresponds to the \emph{first} episode of mass transfer. In contrast, for PTF1\,J2238+7430 the common-envelope phase represents the \emph{second} mass-transfer event, following an earlier phase of stable Roche-lobe overflow. According to \citet{Ge2024}, this distinction may play a key role in explaining the higher efficiency required in the case of PTF1\,J2238+7430. Conversely, \citet{Nelemans2025} found that the first phase of mass transfer in low-mass systems often proceeds neither as stable mass transfer nor as a common-envelope phase. Therefore, if the physics of stable mass transfer is not fully understood, our constraints on $\alpha_{\mathrm{CE}}$ remain uncertain. 
In future work, implementing an explicit treatment of the recombination efficiency in \textsc{MESA}, following for example the approach of \citet{Belloni2025}, could allow us to derive tailored common-envelope efficiencies that depend on the internal structure of each stellar model.

While our simulations successfully identified a configuration that reproduces the observed properties of PTF1,J2238+7430, a major difficulty arose when attempting to obtain sufficiently long post–stable mass-transfer orbital periods ($>400$\,days) consistent with low values of $\alpha_{\mathrm{CE}}$. 
A possible pathway to reach longer orbital periods could involve the enhanced wind prescription recently proposed by \citet{Gao2023}, which increases the orbital separation prior to the onset of mass transfer. Nevertheless, in the case of PTF1\,J2238+7430, this mechanism would only be viable for initially massive main-sequence stars, since the large amounts of mass lost through the wind would otherwise prevent the formation of a sufficiently massive white dwarf ($M_{\mathrm{WD}} \gtrsim 0.7\,M_{\odot}$). This limitation highlights that, under the present assumptions of solar metallicity and suppressed angular momentum accretion, the enhanced wind channel cannot fully reconcile the observed properties of the system. Still, the existence of other long-period sdB+WD binaries with comparable component masses, such as those illustrated in \citet[][their Fig.~8]{Garbutt2024}, suggests that alternative evolutionary pathways or additional physical ingredients may provide multiple viable solutions. Future work should therefore explore these possibilities, including systematic variations in metallicity, angular momentum accretion prescriptions, and mass-loss mechanisms, to better constrain the range of evolutionary channels that can lead to PTF1\,J2238+7430–like systems.

\subsection{Parameter range for similar evolutionary pathways}

\begin{figure}
    \centering
    \includegraphics[width=1.0\linewidth]{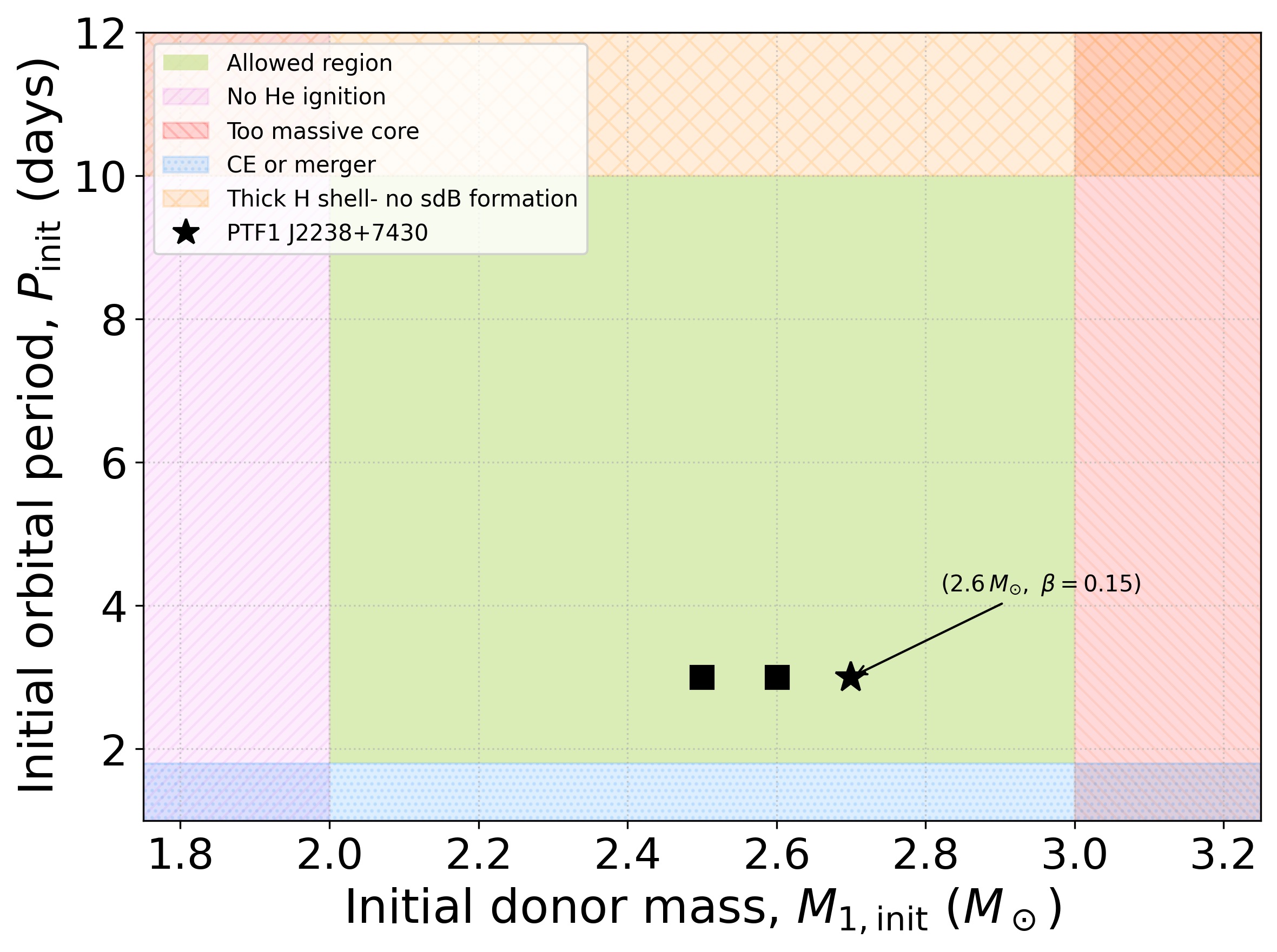}
    \caption{Allowed and excluded regions for the formation of an sdB~+~WD binary through stable mass transfer similar to PTF1~J2238+7430. The central green-shaded region indicates combinations of initial donor mass ($M_{1,\mathrm{init}}$) and orbital period ($P_{\rm init}$) that lead to sdB formation consistent with observations. The hatched and colored regions correspond to physically forbidden configurations: violet (left side) indicates donors too low in mass (no He ignition), red (right side) shows donors too massive (oversized core or Roche-lobe overflow in the subgiant phase), blue (bottom part) denotes systems at risk of common-envelope or merger events, and orange (upper part) represents donors with a thick hydrogen shell preventing sdB formation. The black star marks the parameters of the observed system PTF1~J2238+7430, while black squares represent model systems that produced potential double-detonation supernova progenitors with slightly different masses from the observed system.  }
    \label{fig:limits}
\end{figure}

Although our simulations explore a wide range of parameters  (see appendix Table~\ref{tab:simulaciones}), they were not carried out in a fully systematic manner. Consequently, the trends and limits we identify should be regarded as preliminary indications of the parameter space that appears promising for forming systems similar to PTF1\,J2238+7430, rather than as strict constraints.  

Producing a long-period post–stable mass transfer sdB+WD binary requires a delicate balance between the component masses and the initial orbital period. A mass ratio close to unity ($q \approx 1$) is essential to maintain stability during Roche-lobe overflow, while the donor mass ($M_1$) plays a particularly central role. For $M_1 \gtrsim 3$--$3.5~M_\odot$, the hydrogen-free core mass at the onset of mass transfer tends to exceed the observed sdB mass, implying that sdB formation would require Roche-lobe overflow during the sub-giant phase
\citep{Arancibia-Rojas2024}, which generally produces post--stable mass transfer orbital periods shorter than observed. Overshooting effects in this regime also tend to yield sdB masses larger than those measured in PTF1\,J2238+7430. While more massive sdB stars are not problematic in a general evolutionary context, reproducing the specific properties of PTF1\,J2238+7430 requires forming a relatively low-mass core in the range of $\sim 0.36$--$0.41~M_\odot$. At the other extreme, for $M_1 \lesssim 1.8~M_\odot$, two issues arise. First, helium ignition does not occur at the observed sdB mass in red giants with degenerate cores, preventing the formation of an sdB of the required mass. Second, even for $M_1 \approx 1.8~M_\odot$, the resulting WD progenitor mass after conservative mass transfer is typically too low ($\lesssim 3.6~M_\odot$), which would not allow the formation of a WD with $M \gtrsim 0.7~M_\odot$. Taken together, these trends suggest that viable progenitors are more likely found around $M_1 \sim 2-3~M_\odot$, with roughly equal-mass binaries ($M_1 \approx M_2 \approx 2-3~M_\odot$) appearing particularly favorable for reproducing both the observed mass and orbital period.

The initial orbital period ($P_{\mathrm{orb,init}}$) also shows a strong influence. For periods much longer than $\sim 10$ days, helium ignition may occur before the end of stable mass transfer, preventing sdB formation because a significant fraction of the donor mass has not yet been processed into the helium core. In addition,  in this regime the donor can retain a more massive hydrogen envelope than the typical $\sim  3.16\mathrm{x}10^{-5}-0.01$$\,M_\odot$ found in sdB stars \citep{Bauer2021, Arancibia-Rojas2024, rodriguez2025}, further complicating the reproduction of the observed system.  
On the other hand, very short initial periods (a few days or less) tend to produce post mass-transfer orbits that are overly compact, increasing the risk of Roche-lobe overflow by the accretor, which may drive the system into a common-envelope phase instead of stable mass transfer, or even lead to a merger. In addition, higher $M_1$ values generally require shorter $P_{\rm orb, init}$ to keep the hydrogen-free core mass low, which in turn favors higher mass-loss fractions ($\beta$) to avoid excessive accretor growth. These interdependencies highlight the delicate balance between the different parameters in reproducing the observed system.

The fraction of mass lost from the system during Roche-lobe overflow, denoted as $\beta$, also plays a critical role in the evolution of the binary. Very low values of $\beta$ ($\beta \lesssim 0.15$) lead to excessive accretion onto the secondary, potentially driving it to critical rotation or inducing premature Roche-lobe overflow. In contrast, very high values ($\beta \gtrsim 0.85$--$0.90$) hinder sufficient core growth in the donor before the envelope is removed, preventing helium ignition. Our simulations indicate that only an intermediate range of $\beta$ ($\sim 0.15$--$0.85$) allows for the formation of viable sdB stars.

Figure~\ref{fig:limits} presents a summary highlighting regions of the initial parameter space, particularly the donor mass ($M_{1,\rm init}$) and orbital period ($P_{\rm init}$). The large green-shaded square represents combinations of $M_{1,\rm init}$ and $P_{\rm init}$ that potentially represent initial configurations of post–stable mass-transfer systems leading to the formation of sdB+WD binaries consistent with the observational constraints of PTF1 J2238+7430, 
assuming $q_{\mathrm{crit}} \gtrsim 1$ and $\beta = 0.15$. This parameter space may help to identify systems similar to PTF1\,J2238+7430, such as those marked with black squares, which contain an sdB and a WD with minimum masses of $\sim 0.4\,M_\odot$ and $\sim 0.7\,M_\odot$, respectively. These masses are sufficient to consider them as potential progenitors of double-detonation Type Ia supernovae \citep{Livne1990, Livne1995, Fink2010, Woosley2011, Wang2012,  Wang2018, Shen2014}.
Overall, these constraints provide a useful guide and a foundation for more systematic future explorations of the parameter space.

In summary, while our models identify a plausible parameter space for reproducing the observed system, they rely on a series of tightly constrained assumptions, and our exploration of the initial parameter space remains preliminary and should be regarded as indicative rather than exhaustive. Slight deviations in component masses, initial orbital period, or mass-transfer efficiency can prevent the formation of the observed sdB+WD binary, emphasizing the sensitivity of these evolutionary pathways.
 Element diffusion (i.e., gravitational settling and chemical diffusion) is not included in our models. Since diffusion can modify the predicted values of $\log g$ and $T_{\rm eff}$, its inclusion would likely require a slightly different initial primary mass in order to reproduce the present-day properties of the sdB component \citep{Michaud2007, Ostrowski_2021}.
 We have not varied the stellar metallicity, which could in principle affect the internal structure, core growth, and mass-loss rates. Some of these limitations can be alleviated if we relax the restriction of keeping the common-envelope efficiency $\alpha_{\rm CE}$ strictly around $1/3$ and instead allow it to adopt higher values. Future studies should also consider systems with different sdB and WD masses, which could still qualify as potential double-detonation supernova Ia progenitors. Consequently, while our results highlight promising regions of parameter space, further work is needed to more rigorously map all possible evolutionary channels and quantify their likelihood.


Given the configuration we inferred for PTF1~J2238+7430 at the main-sequence stage (initial donor mass $M_{1,\rm init} \sim 2.7~M_\odot$, companion mass $M_{2,\rm init} \sim 2.7~M_\odot$, and initial orbital period $P_{\rm orb, init} \sim 3$~days), \citet{Lagos2020} and \citet{Lagos-Vilches2024} suggest that systems with similar parameters and evolution pathways are likely to host a distant tertiary companion. This is, main-sequence binaries with short orbital periods ($\lesssim 5$~days) have a high probability of being part of a hierarchical triple system \citep{Tokovinin2006}. 
To investigate this possibility, we searched the \textit{Gaia} DR3 catalog for sources around of PTF1~J2238+7430 that have consistent parallax and proper motion values. No candidate objects were found that could plausibly be associated with the system but a deeper search and/or perhaps high angular resolution observations are required to exclude the hypothesis that PTF1~J2238+7430 is part of a hierarchical triple. 


\section{Summary and Conclusions}

Our simulations successfully reproduce the evolutionary pathway of PTF1\,J2238+7430, as originally proposed by \citet{Kupfer22}. Starting from a close binary on the main sequence, we find that stable Roche-lobe overflow of a $\sim2.7\,M_\odot$ donor produces a stripped helium-burning core that matches the observed sdB properties. The accretor grows in mass, reaching nearly $4\,M_\odot$, and subsequently evolves toward the giant branch. At this stage, unstable mass transfer triggers a common-envelope event, leaving behind a $0.406\,M_\odot$ sdB and a $0.72\,M_\odot$ white dwarf in a compact orbit. Gravitational wave radiation then shrinks the orbit to the present-day configuration, with a cooling age for the WD consistent with observations. By providing an evolutionary model that reproduces the observed parameters, our results provide strong support for the stable mass transfer + common-envelope scenario. Moreover, the derived common-envelope efficiency of $\alpha_{\mathrm{CE}} \approx 0.87$ highlights the importance of envelope physics in shaping compact binaries and possible progenitors of thermonuclear supernovae.

However, the reproduction of this system requires finely tuned assumptions. The outcomes depend sensitively on the adopted values of the mass-loss fraction $\beta$, the treatment of angular momentum accretion, and the efficiency of the common-envelope phase. In particular, we find that explaining PTF1\,J2238+7430 requires a relatively high common-envelope efficiency ($\alpha_{\mathrm{CE}} \approx 0.8$), higher than the canonical values inferred for many other post–common-envelope binaries. This difference may reflect the distinct evolutionary history of this system, in which the common envelope corresponds to a \emph{second} mass-transfer phase rather than the first.  

Our results should therefore be viewed as indicative rather than definitive. The parameter space outlined in Figure~\ref{fig:limits} highlights promising regions for forming systems similar to PTF1\,J2238+7430 and may also point to other sdB+WD binaries with component masses sufficient to qualify as potential double-detonation Type Ia supernova progenitors. Furthermore, some areas that we excluded from our analysis---such as progenitors with main-sequence masses $> 3.0\,M_\odot$---could lead to the formation of more massive sdB stars, which in turn may also represent viable double-detonation supernova progenitors. These regimes should therefore be included in future studies aimed at characterizing the broader population of sdB+WD systems.  
Nonetheless, a more systematic exploration—covering variations in metallicity, angular momentum transport, and mass-loss prescriptions—is necessary to fully map the range of viable evolutionary channels and to assess their relative likelihood within stellar populations. Ultimately, extending this work into population synthesis studies will be key to quantifying the frequency of such systems and evaluating their contribution to the overall Type Ia supernova rate.

\begin{acknowledgements}
This research was supported by Deutsche Forschungsgemeinschaft  (DFG, German Research Foundation) under Germany’s Excellence Strategy - EXC 2121 "Quantum Universe" – 390833306. Co-funded by the European Union (ERC, CompactBINARIES, 101078773). Views and opinions expressed are however those of the author(s) only and do not necessarily reflect those of the European Union or the European Research Council. Neither the European Union nor the granting authority can be held responsible for them. MRS acknowledge financial support from FONDECYT
(grant number 1221059). DB acknowledges support from FONDECYT (grant
number 3220167) and the São Paulo Research Foundation (FAPESP), Brazil,
Process Numbers 2024/03736-2 and 2025/00817-4.
\end{acknowledgements}

%
%

\bibliographystyle{aa_url} 
\bibliography{MSHP} 

\begin{appendix}

\section{Summary of Unsuccessful MESA Model Attempts}

In this section, we present a summary of the parameter combinations that were explored but did not lead to viable models. The table lists the initial values adopted for each attempt and the specific reason why the simulation failed to produce an evolution consistent with our target scenario.

\begin{table}[H]
    \centering
    \caption{Summary of parameter combinations tested and the corresponding reasons for unsuccessful outcomes.}

    \begin{tabular}{c c c c l }
    \hline
M1 & M2 & $\beta$ & $P_{\rm orb}$ (d) & Reason for failure \\
\hline

3.5   & 2.7   & 0.25 & 1.8  & Accretor overfills its Roche lobe; need lower M2 and smaller $\beta$. \\

2.301 & 2.300 & 0.40 & 50   & He flash occurs before the end of stable mass transfer. \\
2.301 & 2.300 & 0.40 & 25   & He flash occurs before the end of stable mass transfer. \\
2.301 & 2.300 & 0.40 & 10   & He flash occurs before the end of stable mass transfer. \\
2.301 & 2.300 & 0.40 & 5    & M2 too low. \\
2.301 & 2.300 & 0.30 & 4    & Accretor overfills its Roche lobe. \\

2.31  & 2.29  & 0.20 & 4    & Critical rotation problem. \\

2.31  & 2.30  & 0.20 & 5    & Critical rotation problem. \\

2.51  & 2.50  & 0.40 & 10   & Critical rotation problem. \\
2.51  & 2.50  & 0.40 & 25   & Critical rotation problem. \\

2.501 & 2.500 & 0.40 & 25   & Critical rotation problem. \\
2.501 & 2.500 & 0.50 & 50   & He flash occurs before the end of stable mass transfer. \\

2.101 & 2.100 & 0.20 & 50   & He flash occurs before the end of stable mass transfer. \\
2.101 & 2.100 & 0.20 & 10   & He flash occurs before the end of stable mass transfer. \\
2.101 & 2.100 & 0.10 & 7    & Mass-transfer rate too high ($\dot M > 10^{-2}$). \\

3.001 & 3.000 & 0.80 & 3    & He flash with envelope too large; sdB mass too high; M2 not massive enough; \\

2.401 & 2.400 & 0.30 & 3    & Critical rotation at M2 $\sim$ 3.60 \\
2.401 & 2.400 & 0.35 & 4    & Accretor overfills its Roche lobe. \\
2.401 & 2.400 & 0.35 & 5    & Accretor overfills its Roche lobe. \\
2.401 & 2.400 & 0.35 & 3    & Critical rotation (M2 $\sim$ 3.56). \\
2.401 & 2.400 & 0.25 & 3    & Critical rotation;  M2 core too small for massive WD. \\

2.501 & 2.500 & 0.25 & 3    & Critical rotation (M2 $\sim$ 3.77); final $P_{\rm orb}$ only 166 d; M2 core too small. \\
2.501 & 2.500 & 0.15 & 3    & Critical rotation (M2 $\sim$ 3.87); sdB mass OK; M2 core too small. \\

2.601 & 2.600 & 0.15 & 3    & Critical rotation (M2 $\sim$ 4.0); sdB $T_{\rm eff}$ too low; M2 core insufficient. \\

2.701 & 2.700 & 0.15 & 3    & Critical rotation (M2 $\sim$ 4.14); sdB parameters almost OK; M2 core still too small. \\

2.701 & 2.700 & 0.15 & 3  & $T_{\rm eff}$ improves but $\log g$ and radius do not; accretor overfills its Roche lobe. \\
2.701 & 2.700 & 0.15 & 3.5  & Accretor overfills its Roche lobe. \\

2.710 & 2.700 & 0.10 & 3 & sdB parameters uncertain; accretor overfills its Roche lobe. \\

        \hline\\
     \end{tabular}
    \label{tab:simulaciones}
\end{table}

\end{appendix}

\end{document}